\documentclass[twocolumn,pra]{revtex4}

\usepackage[T1]{fontenc}

\usepackage{graphicx}
\usepackage{amsmath,amssymb}
\usepackage{bm}
\usepackage{textcomp,gensymb,units}

\newcommand{\vc}[1]{\ensuremath{\bm{#1}}}
\newcommand{\bra}[1]{\ensuremath{\left\langle {#1} \right|}}
\newcommand{\ket}[1]{\ensuremath{\left|  #1 \right\rangle}}

\newcommand{\aver}[1]{\ensuremath{\langle {#1} \rangle}}

\newcommand{\abs}[1]{\ensuremath{\left| {#1} \right|}}

\newcommand{\comm}[2]{\ensuremath{\left[ {#1} , {#2} \right]}}

\newcommand{\Sx}{\ensuremath{S_x}}
\newcommand{\Sy}{\ensuremath{S_y}}
\newcommand{\Syp}{\ensuremath{\tilde{S}_y}}
\newcommand{\Sz}{\ensuremath{S_z}}
\newcommand{\Sp}{\ensuremath{S_+}}
\newcommand{\tSp}{\ensuremath{\tilde{S}_+}}

\newcommand{\DSz}{\ensuremath{\Delta S_z}}

\newcommand{\vSz}{\ensuremath{\Delta S_z^2}}
\newcommand{\vSyp}{\ensuremath{ \Delta \tilde{S}_y^2}}

\newcommand{\vSphi}{\ensuremath{\Delta S_\alpha^2}}

\newcommand{\vScurv}{\ensuremath{\Delta S_\mathrm{curv}^2}}

\newcommand{\vcurv}{\ensuremath{\sigma_\mathrm{curv}^2}}
\newcommand{\vfs}{\ensuremath{\sigma_{\alpha_0,r}^2}}

\newcommand{\one}{\ket{\uparrow}}
\newcommand{\uu}{\ket{\uparrow}_i\bra{\uparrow}_i}
\newcommand{\ud}{\ket{\uparrow}_i\bra{\downarrow}_i}
\newcommand{\dd}{\ket{\downarrow}_i\bra{\downarrow}_i}

\newcommand{\eu}{\ket{e}_i\bra{\uparrow}_i}

\newcommand{\ed}{\ket{e}_i\bra{\downarrow}_i}
\newcommand{\ee}{\ket{e}_i\bra{e}_i}
\newcommand{\two}{\ket{\downarrow}}
\newcommand{\exc}{\ket{e}}

\newcommand{\cd}{c^\dagger}

\newcommand{\raw}{\rightarrow}
\newcommand{\lraw}{\leftrightarrow}

\newcommand{\ddt}[1]{\ensuremath{\frac{d{#1}}{dt}}}

\begin{document}

\title{Squeezing the Collective Spin of a Dilute Atomic Ensemble by Cavity Feedback}

\author{Monika H. Schleier-Smith}
\affiliation{
Department of Physics, MIT-Harvard Center for Ultracold Atoms,
and Research Laboratory of Electronics, Massachusetts Institute of Technology,
Cambridge, Massachusetts 02139, USA}

\author{Ian D. Leroux}
\affiliation{
Department of Physics, MIT-Harvard Center for Ultracold Atoms,
and Research Laboratory of Electronics, Massachusetts Institute of Technology,
Cambridge, Massachusetts 02139, USA}

\author{Vladan Vuleti\'{c}}
\affiliation{
Department of Physics, MIT-Harvard Center for Ultracold Atoms,
and Research Laboratory of Electronics, Massachusetts Institute of Technology,
Cambridge, Massachusetts 02139, USA}

\date{\today}

\begin{abstract}
We propose and analyze a simple method to squeeze dynamically and unconditionally the collective spin of a dilute atomic ensemble by interaction with a driven mode of an optical resonator, as recently demonstrated [I.~D.~L., M.~H.~S., and V.~V., Phys. Rev. Lett. 104, 073602 (2010)].  We show that substantial squeezing can be achieved in the regime of strong collective ensemble-resonator coupling.  The squeezing is ultimately limited either by photon emission into free space or by the curvature of the Bloch sphere.  We derive both limits and show where each prevails.

\end{abstract}

\maketitle

While techniques for preparing single-particle spin states in atomic ensembles (coherent spin states \cite{Arecchi72}) are well established, the manipulation of arbitrary quantum mechanical many-body states remains far out of reach, and only a tiny fraction of a many-spin Hilbert space is experimentally accessible to date. The preparation of even weakly quantum-correlated (entangled) states requires a controllable interaction between the particles \cite{Kitagawa93} that induces system evolution that is fast compared to system decoherence.  Only a few (pseudo-)spin systems offer such favorable interaction-to-decoherence ratios, notably trapped ions, where entangled states of up to eight particles have been prepared \cite{Haffner05,Leibfried05}, and colliding atoms in a Bose-Einstein condensate \cite{Sorensen01a}. In those systems, squeezed spin states \cite{Kitagawa93,Wineland92,Wineland94}---in which the quantum noise is redistributed so that one noise component is smaller than possible for unentangled states---have been generated \cite{Meyer01,Esteve08}.

In a dilute atomic ensemble, a common interaction of the atoms with a light field can replace direct spin-spin interactions \cite{Ueda96,Kuzmich98,Meiser08}.  Entanglement between the spin and light degrees of freedom allows a measurement performed on the light to reduce spin noise \cite{Kuzmich98,Kuzmich00} and produce spin squeezing conditionally, as recently demonstrated by two groups using trapped atoms \cite{Appel09,SchleierSmith10a}.

Pioneering proposals \cite{Zhang03,Hammerer04,Takeuchi05} have shown that repeated light-ensemble interaction can produce spin squeezing dynamically and deterministically, without requiring measurement of the light field. Such processes can be viewed as quantum coherent feedback \cite{Wiseman94b}: the ensemble spin imprints its quantum fluctuations on the light, which acts back on the spin state to reduce those fluctuations.

\begin{figure}
  \includegraphics[width=0.9\columnwidth]{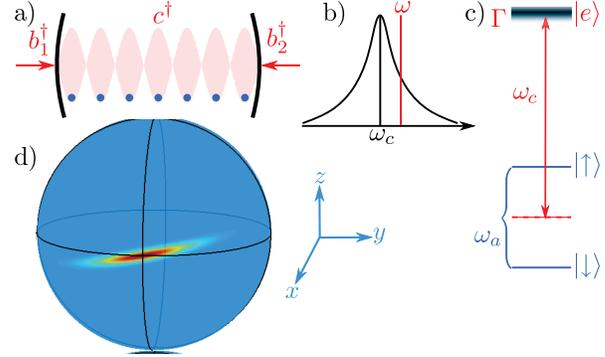}
  \caption{(Color online) Scheme for spin squeezing by cavity feedback.  An ensemble of atoms is uniformly coupled to a laser field in an optical resonator (a).  A laser tuned to the slope of a cavity resonance (b), at equal and opposite detunings from transitions $\ket{\uparrow}\rightarrow\ket{e}$ and $\ket{\downarrow}\rightarrow\ket{e}$ (c), shears a coherent spin state prepared along the $x$ axis into a squeezed spin state (d), illustrated by a tomographic probability distribution \cite{Cunha01} on the Bloch sphere.
  }\label{fig:setup}
\end{figure}

Here we propose a simple and robust method for squeezing the spin of an atomic ensemble inside an optical cavity by coherent feedback.  We show that a laser tuned to the slope of the cavity resonance and off-resonant from the atomic transition induces a one-axis twisting \cite{Kitagawa93,Takeuchi05} of the spin state space (Bloch sphere; Fig. 1d).  For an initial coherent spin state (CSS) \cite{Arecchi72} orthogonal to the twisting axis, this reduces the quantum noise in an oblique spin component.  While photon shot noise in the incident light and photon scattering into free space counteract the noise reduction, substantial squeezing is possible provided the collective cooperativity (resonant optical depth) is large, even if single atoms are weakly coupled to the cavity.


For an ensemble of two-level (spin-$\frac{1}{2}$) atoms, described by a collective spin $\vc{S}$, the squeezing is produced by an ensemble-light interaction Hamiltonian of the form $\cd c \Sz$ that represents the differential energy shift between the two atomic states due to the intracavity light with photon number $\cd c$.  This photon number depends on the population difference $2\Sz$ between the atomic states because the precise tuning of the resonator mode relative to the driving laser depends on the atoms' state-dependent index of refraction.  In particular, if $\cd c$ depends linearly on $\Sz$, the differential energy shift causes a precession of the spin vector about the $z$ axis at a rate proportional to $\Sz$, similar to the dynamics of a Hamiltonian $\propto\Sz^2$ \cite{Kitagawa93}. The initially circular uncertainty region of a CSS is then sheared into an ellipse (Fig. \ref{fig:setup}d) that is narrower in one direction than the original CSS, corresponding to spin squeezing. The spin correlations between different atoms arise from the fact that the phase between the two states in any individual atom now depends on the state population difference $2 \Sz$ of the entire ensemble.

We consider an ensemble of $N$ identical three-level atoms whose ground states $\one,\two$ (e.g. hyperfine states) are split by an energy $\hbar \omega_a$ and coupled via optical transitions of frequencies $\omega_c \pm \omega_a/2$ to an excited state $\exc$ with population decay rate $\Gamma$ (see Fig.~\ref{fig:setup}c).  We assume that a resonator mode of interest, with linewidth $\kappa$, is tuned to the frequency $\omega_c$, i.e. it has equal and opposite detunings $\pm\Delta = \pm \omega_a/2$ from the transitions $\one \lraw \exc, \two \lraw \exc$. For simplicity, we assume that the two transitions have equal strength and that all atoms are equally coupled to the resonator, with single-photon Rabi frequency $2g$.  The Hamiltonian is then given by 
\begin{align}\label{eq:fullH}
H = \hbar\omega_c \cd c + &\hbar \sum_{i=1}^N \Big{(}\frac{\omega_a}{2}\left[\uu-\dd\right] +\omega_c\ee \nonumber\\&+ g\left[c\eu + c\ed + H.c.\right]\Big{)},
\end{align}
where $c,\cd$ are the photon annihilation and creation operators for the cavity mode and the index $i$ labels the individual atoms.

As we are interested in effects of the light shift and not in populating the excited state, we assume a large detuning $|\Delta| \gg \kappa, \Gamma, g$.  For sufficiently low intracavity photon number $\aver{\cd c}\ll (\Delta/g)^2$, we can adiabatically eliminate the excited state \cite{Brion07} to arrive at an effective Hamiltonian for the dynamics in the two ground states:
\begin{equation}\label{eq:Hamiltonian}
    H_\mathrm{eff} = \hbar \omega_c \cd c +  \hbar \Omega \cd c \Sz + \hbar \omega_a \Sz,
\end{equation}
where $\Sz = \sum_{i=1}^{N}(\uu-\dd)/2$ and $\Omega = 2 g^2/|\Delta|$.  (We shall later treat semiclassically the effect of spontaneous emission from $\exc$ into free space.)  The first term in $H_\mathrm{eff}$ describes the energy of the free field in the cavity, while the last term represents the energy of the atomic system with population difference $2 \Sz$ between the two states. The interaction term $\hbar \Omega \cd c \Sz$ can be alternatively grouped with the first term to describe the shift of the cavity resonance by an amount $\Omega \Sz$ due to the atomic-state-dependent index of refraction originating from the transitions $\one \lraw \exc$ and $\two \lraw \exc$, or with the last term to describe the light shifts $\pm \hbar \Omega \cd c /2$ experienced by the atoms in states $\one$ and $\two$.

We adopt an interaction picture where the spin state vector evolves with the atomic Hamiltonian $H_a=\hbar \omega_a \Sz$ and the spin operator $\vc{S}$ evolves with the Hamiltonian $H_0 = \hbar \left(\omega_c  +  \Omega \Sz \right) \cd c$. Causality requires $\comm{c^{(\dagger)}(t)}{\vc{S}(t)}=0$ \cite{Gardiner85,WallsMilburn}.  It follows that $\Sz$ is a constant of motion, while $\Sp = \sum_{i=1}^N\ud=\Sx + i\Sy$, which characterizes the phase on the Bloch sphere, evolves as
\begin{equation}
\label{eq:SpMotion}
  \ddt{S_+^n} = i \hbar^{-1} \comm{H_0}{\Sp^n} = i n\Omega \cd\Sp^n c
\end{equation}
for any positive integer $n$.

The equation of motion for the cavity annihilation operator $c$ arises from the interaction both with the atomic system, through $H_0$, and with the bath modes outside the cavity, and can be derived with the standard input-output formalism for cavity fields \cite{Gardiner85,WallsMilburn}:
\begin{equation}
\label{eq:aMotion}
  \dot{c} = \left( - \frac{\kappa}{2} -i \omega_c  - i \Omega \Sz \right) c + \sqrt{\frac{\kappa}{2}}(b_1+b_2),
\end{equation}
where $b_1,b_2$ are the annihilation operators for the input fields from the two sides of the cavity (Fig. \ref{fig:setup}). The term $i \Omega \Sz$ describes the atomic tuning of the cavity frequency.


We consider a monochromatic driving field, input from one side, in a coherent state $\ket{\beta}_{\omega}$ defined by $b_1 (t) \ket{\beta}_{\omega}= \sqrt{\kappa} e^{-i\omega t} \beta \ket{\beta}_{\omega}$ and $b_2(t)\ket{\beta}_{\omega}=0$, where the frequency $\omega=\omega_c+\kappa/2$ is tuned to the slope of the resonator mode.  In order to resolve the cavity resonance, we consider interaction times $t\gg\kappa^{-1}$, and we assume that the field is turned on adiabatically so that its transient behavior can be neglected.  We consider an atomic system prepared initially in a CSS along the $x$ axis \cite{Arecchi72}, satisfying $\Sx(0) \ket{\psi_0} = S \ket{\psi_0}$.  To obtain a linear shearing of the CSS uncertainty region, we assume $\Omega \sqrt{S/2}\ll\kappa$ so that the fluctuations of $S_z$ with variance $\DSz^2=S/2$ induce proportional fluctuations in intracavity power.

The steady-state solution to Eq. \ref{eq:aMotion} yields $c(t)\ket{\beta}_{\omega} = e^{-i\omega t}\kappa\beta/(\sqrt{2}(\gamma-i\omega))\ket{\beta}_{\omega}$, where the complex frequency $\gamma=\kappa/2 + i( \omega_c+\Omega \Sz)$ accounts for the leakage through the cavity mirrors as well as the phase evolution of the field.  We substitute this result into Eq. \ref{eq:SpMotion} to obtain the time evolution of arbitrary powers of $S_+$ in the spin subsystem,
\begin{equation}\label{eq:SpSolDiff}
\ddt{\aver{\Sp^n}_\beta} = i f_n(\Sz)\aver{\Sp^n}_\beta,
\end{equation}
where $\aver{\hat{\mathcal{O}}}_\beta\equiv\bra{\beta}_\omega\hat{\mathcal{O}}\ket{\beta}_\omega$ denotes a trace over the field state.  
To lowest order in $(\Omega/\kappa)\abs{\Sz}\lesssim (\Omega/\kappa)\sqrt{S/2}\ll 1$ and for $n\lesssim\sqrt{S}$,
\begin{equation}\label{eq:fn}
f_n(\Sz) = n\Omega\abs{\beta}^2(1+n(i-1)\Omega/\kappa + 2(\Omega/\kappa)\Sz).
\end{equation}
For easier visualization, after application of the squeezing light the spin is quickly rotated back about $z$, $\tilde{S}_+ \equiv \Sp(t) e^{- i f_1(0)t}$, by the angle $f_1(0)t$ corresponding to the light level for $\Sz=0$.  It is useful to introduce the dimensionless shearing strength $Q=S p_0 (2 \Omega/\kappa)^2$, proportional to the spin $S$, average photon number $p_0=|\beta|^2 \kappa t/2$ transmitted for $\Sz=0$ in the time $t$, and square of the differential atomic phase shift per transmitted photon $2\Omega/\kappa \ll 1$.  Equations \ref{eq:SpSolDiff}--\ref{eq:fn} then yield, in terms of the initial spin operators $\Sp(0)$ and $\Sz$,
\begin{eqnarray}\label{eq:Sp}
\aver{\tSp}_\beta &=& e^{i Q \Sz/S}\Sp(0)\\
\label{eq:SpSq}
\aver{\tSp^2}_\beta &=& e^{-(1+i)Q/S}e^{2iQ\Sz/S}\Sp^2(0).
\end{eqnarray}

Using Eqs. \ref{eq:Sp}--\ref{eq:SpSq} and Ref. \cite{Arecchi72}, we can evaluate various spin expectation values of interest for the atomic input state $\ket{\psi_0}$. In particular, 
\begin{align} \label{eq:varY1}
  \vSyp = \frac{S^2}{2}+\frac{S}{4} -\left(\frac{S^2}{2}- \frac{S}{4} \right)e^{-Q/S}\mathcal{G}_S(Q),\\
   \label{eq:covYZ1}
  \aver{\Syp \Sz + \Sz \Syp} = (2S^2-S)\sin\left(\frac{Q}{2S}\right)\mathcal{G}_S(Q/2),
\end{align}
and $\vSz = S/2$, where $\mathcal{G}_S(u)\equiv\cos^{2S-1}(u/S)$.  The correlation term $\aver{\Syp \Sz + \Sz \Syp}$, which is zero for a CSS (Q=0), displays the quantum correlations between $\Syp$ and $\Sz$ induced by the $\Sz$-dependent spin precession.  In  the limit of a large ensemble, $S\gg 1$, and small phase $|Q \Sz/S| \sim Q/\sqrt{S}\ll 1$, we have
\begin{equation} \label{eq:varY2}
  \vSyp \approx \frac{S}{2}\left(1+Q+Q^2\right).
\end{equation}
The three terms in this expression represent the variance $S/2$ of the initial CSS, the variance increase due to photon shot noise $Q S/2 \propto p_0 \propto t$, and the variance increase due to cavity-mediated coherent feedback $Q^2 S/2 \propto p_0^2 \propto t^2$. Thus, for $Q>1$ feedback stretches the uncertainty region more quickly than does photon shot noise. This allows spin squeezing by one-axis twisting in the open quantum system, even though tracing over the output light field results in dissipative dynamics of the spin subsystem.

To verify that the spin state is squeezed, we calculate the spin variance along the $z$ axis after rotation about the $x$ axis by an angle $-\alpha$:
\begin{equation}\label{eq:Sbeta}
    \vSphi = \frac{1}{2} \left(V_+ - \sqrt{V_-^2+W^2} \cos \left[2(\alpha - \alpha_0)\right]  \right),
\end{equation}
where $V_\pm=\vSyp \pm \vSz$, $W=\aver{\Syp \Sz + \Sz \Syp}$, and $\tan{2\alpha_0}=W/V_-$ are specified by Eqs. \ref{eq:varY1}-\ref{eq:covYZ1}.

For moderate squeezing $1 \ll Q^2 \ll S$, the minimum and maximum variances, normalized to the CSS variance, are $\sigma_{\alpha_0}^2 \approx 1/Q$ and $\sigma_{\alpha_0+\pi/2}^2 \approx Q^2$, where $\sigma_{\alpha}^2\equiv\vSphi/(S/2)$.  While the antisqueezing increases as $Q^2$ due to cavity feedback (Eq. \ref{eq:varY2}), the minimum variance decreases only as $Q^{-1}$ because of the uncertainty in spin precession angle resulting from photon shot noise. The growth in the uncertainty product $\sigma_{\alpha_0}\sigma_{\alpha_0+\pi/2}\approx\sqrt{Q}$ results from ignoring information in the outgoing light, which is entangled with the ensemble spin.

For a given atom number $2S$, the squeezing improves with photon number $p_0$ until the curvature of the Bloch sphere leads to reduced correlation between $\Syp$ and $\Sz$ \cite{Kitagawa93,Takeuchi05}. Expanding $\Delta S_{\alpha_0}^2$ to lowest order in the characteristic phase variance $Q^2/(2S)$, we find $\sigma_{\alpha_0}^2=Q^{-1}+Q^4/(24 S^2)$. The curvature of the Bloch sphere thus limits the minimum uncertainty to $\vcurv = (5/4)6^{-1/5}S^{-2/5}$, reached at a shearing strength $Q_\mathrm{curv}=6^{1/5} S^{2/5}$.  This is the same scaling with atom number as derived for a related scheme in free space \cite{Takeuchi05}.

The attainable squeezing is also limited by photon emission into modes other than the mode of interest \cite{Madsen04,Hammerer04,Saffman09}.  Squeezing requires two-fold light-atom interaction, namely the tuning of the cavity by the atoms in combination with the light shift by the modified intracavity intensity.  As each process is proportional to the real part  $\mathrm{Re}(\alpha) \propto \Omega$ of the atomic polarizability $\alpha$, the shearing strength $Q \propto \Omega^2 \propto \mathrm{Re}(\alpha)^2 \propto \Delta^{-2}\propto\mathrm{Im}(\alpha)$ is proportional to the average number of photons $2r$ emitted into free space per atom during the squeezing process.  Therefore, free-space scattering cannot be ignored at any light-atom detuning $\Delta$, as is also evident if we express the shearing strength $Q$ in terms of the single-atom cooperativity $\eta=4 g^2/(\kappa \Gamma)$ as $Q=4 S \eta r$.  This expression furthermore shows that achieving $Q>1$ at low photon scattering probability, $r \ll 1$---as is necessary to maintain coherence in the system---requires a large collective cooperativity, $S \eta \gg 1$.

\begin{figure}
  \includegraphics[width=0.9\columnwidth]{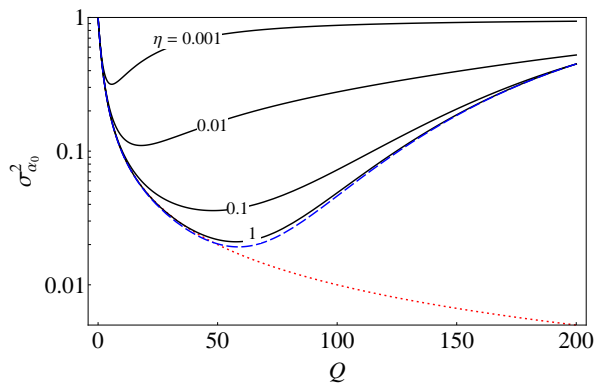}
  \caption{(Color online) Minimum normalized variance $\vfs$ as a function of shearing strength $Q$ for $S=10^4$ and various single-atom cooperativities $\eta = 0.001, 0.01, 0.1, 1$ (solid lines). The dashed line shows the limit $\vcurv$ due to the curvature of the Bloch sphere when free-space scattering is ignored.  The dotted line shows the variance neglecting both free-space scattering and curvature, scaling as 1/Q for $Q\gg 1$.
  }\label{fig:squeeze}
\end{figure}

For the symmetric level scheme considered here (Fig. \ref{fig:setup}), Rayleigh scattering (where the atom returns to the same ground state) occurs at equal rates for states $\one,\two$ and causes no decoherence since it provides no information about the atomic spin \cite{Ozeri05}. The random changes in $\Sz$ arising from Raman scattering $\one \raw \exc \raw \two$ or $\two \raw \exc \raw \one$, on the other hand, reduce the correlation between the time average $\overline{S}_z$ during the squeezing, which determines the evolution of $S_y$, and the final value ${\Sz}(t)$ when the rotation about the $x$-axis is performed.  Replacing $\Sz$ by $\overline{\Sz}$ in Eqs. \ref{eq:Sp} and \ref{eq:SpSq} and evaluating the modified variance $\vSyp$ and covariance $\aver{\Syp \Sz(t) + \Sz(t) \Syp}$ in the large-$S$ limit using $2\aver{\Sz(t_1)\Sz(t_2)}/S=e^{-2r|t_1-t_2|/t}$, we apply Eq. \ref{eq:Sbeta} to calculate the minimum normalized variance $\vfs$.  To lowest order in the number $r\ll 1$ of Raman-scattered photons, and ignoring curvature effects for the moment,
\begin{equation}\label{VarScattering}
    \vfs \approx \left( \frac{1}{Q} + \frac{Q}{3S \eta}  \right) =  \left( \frac{1}{4 S \eta r} + \frac{4 r}{3}  \right).
\end{equation}
As a function of $r$ or time, the variance first decreases below the CSS variance due to the coherent feedback but eventually rises again when the noise added by photon scattering into free space becomes appreciable (Fig. \ref{fig:squeeze}). Consequently, there is an optimum shearing strength $Q_\mathrm{scatt}=\sqrt{3S\eta}$, at which the dynamic squeezing reduces the spin variance by a factor $1/\vfs=\sqrt{3S\eta}/2$ below that of the CSS.  (This is the same scaling as in measurement-induced squeezing \cite{Madsen04,Hammerer04}.)  Squeezing is thus possible even for a resonator that is weakly coupled to a single atom ($\eta \ll 1$), as long as the collective cooperativity is large ($S \eta \gg 1$). The limit $\vScurv$ is only reached for $S\eta^5\gtrsim 1$; this is readily satisfied in moderately coupled resonators ($\eta \gtrsim 1$), but in free-space-like \cite{Takeuchi05} situations ($\eta \ll 1$), for any reasonable atom number $2S$, the squeezing is limited by scattering long before the curvature of the Bloch sphere becomes significant.

By projecting atoms into $\one$ and $\two$, Raman scattering not only adds noise but also shortens the spin vector.  Provided $S \eta \gg 1$, however, this effect is negligible at the optimum squeezing point $r_\mathrm{opt}=\sqrt{3/(16 S \eta)} \ll 1$.  Note also that it may be possible to find other transition schemes where Raman scattering is suppressed \cite{Saffman09}.

Fig. \ref{fig:squeeze} shows the minimum variance $\vfs$ as a function of shearing strength $Q$ for various values of the cooperativity $\eta$, calculated from the full expressions of Eqs.~\ref{eq:varY1}--\ref{eq:covYZ1} modified as described above to account for Raman scattering.  The parameters $\eta=0.1$, $S=10^4$ are similar to those used to achieve squeezing in Refs. \cite{SchleierSmith10a,Leroux10}.

The three approximations of low saturation of the optical transitions, adiabaticity of the input pulse on the scale of the cavity lifetime, and small tuning of the cavity by the atomic quantum noise are all realistic.  Since the excited-state population $\epsilon=\aver{\cd c}g^2/\Delta^2$ is inversely proportional to the interaction time $t$, with $\epsilon \kappa t =(\kappa/g)^2 Q/(8S)$, the first two conditions $\epsilon\ll 1$ and $\kappa t\gg 1$ are always consistent.  For $S=10^4$, $\eta = 0.1$, $\kappa = 2\pi\times 1$~MHz, $g=2\pi\times 0.4$~MHz, and $\Delta/\Gamma = 500$ \cite{SchleierSmith10a,Leroux10}, requiring e.g. $\epsilon \le 10^{-5}$ also ensures that the optimum shearing parameter $Q \approx 50$ is only reached at $t \gtrsim 400/\kappa$.  The dependence of intracavity power on $S_z$ stays well in the linear regime, as $\Omega\sqrt{S/2} = 7\times 10^{-3}\kappa$.

The expressions derived here well describe recent results using a dilute ensemble of $^{87}$Rb atoms inside an optical resonator in the moderate-coupling regime \cite{Leroux10}. Unlike measurement-induced squeezing \cite{Appel09,SchleierSmith10a}, the present method does not require detection of the squeezing light and produces unconditionally squeezed states.  Such states may be used in an atomic clock \cite{Santarelli99,Ye08} to reduce the quantum projection noise of the readout. Our scheme is applicable to both microwave and optical clock transitions, as it can be generalized to any configuration with different light shifts for the two ground states.


This work was supported in part by the NSF, DARPA, and the NSF Center for Ultracold Atoms.  M.~S. acknowledges support from the Hertz Foundation and NSF. I.~D.~L.  acknowledges support from NSERC.

\end{document}